\documentclass[11pt,a4paper]{article}
\pdfoutput=1

\usepackage{jcappub}
\usepackage{rotating}
\usepackage{graphicx,epsfig}
\usepackage{amsmath}
\usepackage{amssymb}
\usepackage{subfigure}

\usepackage{relsize}

\def\beq{\begin{equation}}
\def\eeq{\end{equation}}
\def\bea{\begin{eqnarray}}
\def\eea{\end{eqnarray}}

\begin{document}

\title{Structure formation in $f(T)$ gravity and a solution for $H_0$ tension}

\author[a]{Rafael C. Nunes}
\textbf{}
\affiliation[a]{Departamento de F\'isica, Universidade Federal de Juiz de Fora,
36036-330, Juiz de Fora, MG, Brazil}

\emailAdd{rafadcnunes@gmail.com}

\abstract{We investigate the evolution of scalar perturbations in $f(T)$ teleparallel gravity and its effects on the cosmic microwave background 
(CMB) anisotropy. The $f(T)$ gravity generalizes the teleparallel gravity which is formulated on the Weitzenb\"ock spacetime, 
characterized by the vanishing curvature tensor
(absolute parallelism) and the non-vanishing torsion tensor. For the first time, we derive the observational constraints on the modified teleparallel 
gravity using the CMB temperature power spectrum from 
Planck's estimation, in addition to data from baryonic acoustic oscillations (BAO) and local Hubble constant measurements. 
We find that a small deviation of the $f(T)$ gravity model from the $\Lambda$CDM cosmology is slightly favored. 
Besides that, the $f(T)$ gravity model does not show tension on the Hubble 
constant that prevails in the $\Lambda$CDM cosmology.
It is clear that $f(T)$ gravity is also consistent with the CMB observations, and undoubtedly it can serve 
as a viable candidate amongst other modified gravity theories.}


\maketitle

\section{Introduction}

Several extensions of general relativity have been proposed (see \cite{Clifton,Capozziello:2011et,Oikonomou} for a review) and exhaustively investigated to 
explain the observational  data in cosmology and astrophysics. In particular, the additional gravitational degree(s) of freedom from the modified gravity
models are motivated to drive the accelerating expansion of the Universe at late times, as well as at early times (inflation).
Most of the works in this direction start usually from the standard gravitational description, i.e. from its curvature formulation, and extend the
Einstein-Hilbert action in various ways, for instance, the  $f(R)$ gravity \cite{DeFelice:2010aj}, Gauss-Bonnet gravity 
\cite{Nojiri:2005jg}, Lovelock gravity \cite{Lovelock:1971yv}, Ho\v{r}ava-Lifshitz gravity
\cite{Horava:2008ih}, massive gravity \cite{deRham:2014zqa} and several others.
However, one can equally construct the gravitational modifications starting from the torsion-based formulation, and specifically from the 
Teleparallel Equivalent of General  Relativity (TEGR) \cite{Unzicker:2005in,TEGR,TEGR22,Hayashi:1979qx,JGPereira,Maluf:2013gaa}. Since in this
theory the Lagrangian is the torsion scalar $T$, the simplest modification is the $f(T)$ gravity \cite{Bengochea:2008gz,Linder:2010py} 
(see \cite{Cai:2015emx} for a review).


Construction of viable modified teleparallel gravity models has proven to be an efficient candidate other than general relativity and in the last couple 
of years, the cosmological applications of this particular theory have gained a lot of interest in the literature. The accelerated expansions of the universe at both early \cite{Ferraro:2006jd,Bamba:2014zra} and late times \cite{late_time_fT01,late_time_fT02,late_time_fT03,late_time_fT04,late_time_fT05,
late_time_fT06,late_time_fT07} are the outcomes of this theory. The observational data from various astronomical sources indicate 
that the $f(T)$ gravity is a viable alternative to the $\Lambda$CDM-cosmology
\cite{constrains01,constrains02,constrains03,constrains04,constrains05,constrains06,constrains07,constrains08,constrains09,constrains10}, and moreover,
it has been found to be consistent with the solar system constraints \cite{Iorio:2012cm, Farrugia:2016xcw} as well. Additionally, 
there have been some important developments in this direction, see  
\cite{outros01,outros02,outros03,outros04,outros05,outros06,outros07,outros08,outros09,outros10,outros11,outros12,outros13,
outros14,outros15,outros16,outros17,outros18}. The nonlocal deformations of teleparallel gravity have also been developed 
\cite{nonlocal01,nonlocal02,nonlocal03}. However, a long term issue associated with $f(T)$ gravity is that it is not invariant under 
local Loretz tranformations \cite{LSS_fT}, but probably such problem can also be 
solved if one introduces the spin connection along with the tetrad formalisms in $f(T)$ gravity \cite{outros07}. We note that such a 
proposal is under investigation. 

In this work, our aim is to describe the evolution of scalar linear perturbations in $f(T)$ gravity and investigate its effects 
on the cosmic microwave background (CMB) anisotropies. In addition, for the first time, we report the observational constrains on 
modified teleparallel gravity models using the CMB temperature and polarization data from Planck (i.e. Planck TT,TE,EE+low P+lensing reconstruction),
in addition to data from baryonic acoustic oscillations (BAO) and local Hubble constant measurements. 
The manuscript is organized  as follows: In Section \ref{fTcosmology}, we briefly review the  $f(T)$ gravity and its cosmology, 
where the background and perturbative evolutions are given and the effects on the CMB power spectrum are discussed.
In Section \ref{results}, we present the results of observational analysis. Finally, in Section \ref{Conclusions} we summarize our results and 
perspectives. As usual, a subindex zero attached to any quantity means that it must be evaluated at present time. Also,
prime and dot denote the derivatives with respect to the conformal time and cosmic time, respectively.

\section{$f(T)$ gravity and cosmology}
\label{fTcosmology}

In this section, we briefly review $f(T)$ gravity and we apply it in a cosmological framework. 

\subsection{$f(T)$ gravity}

In $f(T)$ gravity, and similarly to all torsional formulations, we use the
vierbein fields $e^\mu_A$, which form an orthonormal base on the tangent space
at each manifold point $x^{\mu}$. The metric then reads as
$g_{\mu\nu}=\eta_{A B} e^A_\mu e^B_\nu$ (in this manuscript greek indices
and Latin indices span respectively the coordinate and tangent spaces). Moreover,
instead of the torsionless Levi-Civita connection, we use the curvatureless
Weitzenb{\"{o}}ck one, $\overset{\mathbf{w}}{\Gamma}^\lambda_{\nu\mu}\equiv e^\lambda_A\:
\partial_\mu e^A_\nu$  \cite{JGPereira}, and hence the gravitational field is
described by the torsion tensor
\begin{equation}
T^\rho_{\verb| |\mu\nu} \equiv e^\rho_A
\left( \partial_\mu e^A_\nu - \partial_\nu e^A_\mu \right).
\end{equation}
The Lagrangian of teleparallel equivalent of general relativity, i.e., the torsion scalar
$T$, is
constructed by contractions of the torsion tensor as \cite{JGPereira}
\begin{equation}
\label{Tscalar}
T\equiv\frac{1}{4}
T^{\rho \mu \nu}
T_{\rho \mu \nu}
+\frac{1}{2}T^{\rho \mu \nu }T_{\nu \mu\rho}
-T_{\rho \mu}{}^{\rho }T^{\nu\mu}{}_{\nu}.
\end{equation}
Inspired by the $f(R)$ extensions of general relativity,  we can  extend $T$ to a function
$f(T)$, constructing the action of $f(T)$ gravity \cite{Bengochea:2008gz,Linder:2010py} as
\begin{eqnarray}
\label{actionbasic}
 {\mathcal S} = \frac{1}{16\pi G}\int d^4x e \left[f(T)\right],
\end{eqnarray}
with $e = \text{det}(e_{\mu}^A) = \sqrt{-g}$ and $G$ the gravitational
constant. where we have imposed units where the light speed is equal to 1. Note that
TEGR and thus the general
relativity is restored when $f(T)=T$, whereas we recover general relativity with a cosmological constant for $f(T)= T + \Lambda$.

\subsection{$f(T)$ cosmology}

In what follows, we describe the general formalism/equations of the background and scalar perturbation evolution for modified teleparallel cosmology. 

\subsubsection{Background evolution}

We apply $f(T)$ gravity in a cosmological framework, considering the functional form: $f(T)= T + F(T)$. Firstly, we need to incorporate
the matter (baryons and cold dark matter) and the radiation (photons and neutrinos) sectors, and thus the total
action is written as
\begin{eqnarray}
\label{action11mat}
 {\mathcal S}_{tot} = \frac{1}{16\pi G }\int d^4x e
\left[T+F(T)\right] \, + {\mathcal S}_m+ {\mathcal S}_r   ,
\end{eqnarray}
with the matter and radiation Lagrangians  assumed to correspond
to perfect fluids with energy densities $\rho_m$, $\rho_r$ and pressures $P_m$, $P_r$
respectively.

Variation of the action (\ref{action11mat}) with
respect to the vierbeins provides the field equations as
\begin{eqnarray}
\label{eom}
&&\!\!\!\!\!\!\!\!\!\!\!\!\!\!\!
e^{-1}\partial_{\mu}(ee_A^{\rho}S_{\rho}{}^{\mu\nu})[1+F_{T}]
 +
e_A^{\rho}S_{\rho}{}^{\mu\nu}\partial_{\mu}({T})F_{TT}
-[1+F_{T}]e_{A}^{\lambda}T^{\rho}{}_{\mu\lambda}S_{\rho}{}^{\nu\mu}+\frac{1}{4} e_ { A
} ^ {
\nu
}[T+F({T})] \nonumber \\
&&= 4\pi Ge_{A}^{\rho}
\left[{\mathcal{T}^{(m)}}_{\rho}{}^{\nu}+{\mathcal{T}^{(r)}}_{\rho}{}^{\nu}\right],
\end{eqnarray}
with $F_{T}=\partial F/\partial T$, $F_{TT}=\partial^{2} F/\partial T^{2}$,
and where ${\mathcal{T}^{(m)}}_{\rho}{}^{\nu}$ and  ${\mathcal{T}^{(r)}}_{\rho}{}^{\nu}$
are  the matter and radiation energy-momentum tensors respectively.

As a next step, we focus on homogeneous and isotropic geometry, considering the
usual choice for the vierbiens, namely
\begin{equation}
\label{weproudlyuse}
e_{\mu}^A={\rm
diag}(1,a,a,a),
\end{equation}
which corresponds to a flat Friedmann-Robertson-Walker (FRW) background metric.


Inserting the vierbein   (\ref{weproudlyuse}) into the field equations
(\ref{eom}), we acquire the Friedmann equations as
\begin{eqnarray}\label{background1}
&&H^2= \frac{8\pi G}{3}(\rho_m+\rho_r)
-\frac{f}{6}+\frac{TF_T}{3},\\\label{background2}
&&\dot{H}=-\frac{4\pi G(\rho_m+P_m+\rho_r+P_r)}{1+F_{T}+2TF_{TT}},
\end{eqnarray}
with $H\equiv\dot{a}/a$ the Hubble parameter, and where we use dots to denote derivatives
with respect to cosmic time $t$. In the above relations, we have used 
\begin{eqnarray}
\label{TH2}
T=-6H^2,
\end{eqnarray}
which  arises straightforwardly for a FRW
universe through (\ref{Tscalar}).

Observing the form of the first Friedmann equation (\ref{background1}), we deduce that
in $f(T)$ cosmology  we acquire  an effective dark energy sector of
gravitational origin. In particular, we can define the effective
dark energy density as \cite{Cai:2015emx}

\begin{eqnarray}
 \rho_{DE}\equiv\frac{3}{8 \pi G} \left[-\frac{f}{6}+\frac{TF_T}{3} \right].
\end{eqnarray}


In what follows, a subindex zero attached to any quantity implies its value at the present time. In this work, we are interested in confronting
the model with observational data. Hence, we firstly define 
\begin{eqnarray}
\label{THdef3}
\frac{H^2(z)}{H^2_{0}}=\frac{T(z)}{T_{0}},
\end{eqnarray}
with $T_0\equiv-6H_{0}^{2}$.

%

Therefore, using additionally that
$\rho_{m}=\rho_{m0}(1+z)^{3}$, $\rho_{r}=\rho_{r0}(1+z)^{4}$, we
re-write the first Friedmann equation (\ref{background1}) as \cite{constrains01,constrains02}
\begin{eqnarray}
\label{Mod1Ez}
\frac{H^2(z,{\bf r})}{H^2_{0}}=\Omega_{m0}(1+z)^3+\Omega_{r0}(1+z)^4+\Omega_{F0} y(z,{\bf r})
\end{eqnarray}
where
\begin{equation}
\label{LL}
\Omega_{F0}=1-\Omega_{m0}-\Omega_{r0} \;,
\end{equation}
with $\Omega_{i0}=\frac{8\pi G \rho_{i0}}{3H_0^2}$ the corresponding
density parameter at present. In this case the effect of the $F(T)$ modification  is
encoded in the function  $y(z,{\bf r})$ (normalized to
unity at   present time), which depends on $\Omega_{m0},\Omega_{r0}$, and on the
$F(T)$-form parameters $r_1,r_2,...$, namely \cite{constrains01,constrains02}:
\begin{equation}
\label{distortparam}
 y(z,{\bf r})=\frac{1}{T_0\Omega_{F0}}\left[f-2Tf_T\right].
\end{equation}

We mention that due to  (\ref{TH2}), the additional term (\ref{distortparam})
in the effective Friedman equation (\ref{Mod1Ez})   is a function of the
Hubble parameter only.

\subsubsection{Scalar perturbation in $f(T)$ gravity}

In this subsection, we describe how the linear scalar perturbations evolve in the context of $f(T)$ gravity, and quantify its effects
on the CMB anisotropies. Here, we follow the methodology used in \cite{LSS_fT}. The evolution of the matter density perturbations 
in modified teleparallel gravity theories has also been investigated in \cite{Chen:2010va,LSS_fT01}.

We adopt the conformal Newtonian gauge, where the line element of the linearly perturbed FLRW metric is given by
 
\begin{equation}
\label{metric_N}
 ds^2 = a^2(\tau)[-(1+2\psi)d\tau^2 + (1 - 2\phi)\gamma_{ij}dx^idx^j].
\end{equation}
Here $\tau$ is the conformal time, $\psi$ and $\phi$ are the Bardeen potentials, and $\gamma_{ij}dx^idx^j$ is the spatial part of the metric.

The above metric can be mapped considering a decomposition of the vierbein as $e_{\mu}^A = \overline{e}_{\mu}^A + \zeta_{\mu}^A$, 
where $\zeta_{\mu}^A$ is a purely perturbed quantity. The vierbein decomposition satisfies 
$g_{\mu \nu} = \eta_{A B} \overline{e}_{\mu}^A  \overline{e}_{\mu}^B$. Thus, the perturbation of $\overline{e}_{\mu}^A$ 
has the most general form given by

\begin{equation}
\begin{array}{r@{}l}
\overline{e}_{\mu}^0 = \delta_{\mu}^0(1+\psi) + a \delta_{\mu}^i(G_i + \partial_i F),  \\ 
\overline{e}_{\mu}^a = a\delta_{\mu}^a(1-\phi) + a \delta_{\mu}^i(h_{\mu}^i + \partial_i \partial^a B + \partial^a C_i),
\end{array}
\end{equation}
where $C_i$ and $G_i$, $h_{ij}$ are the transverse vector modes and traceless tensor modes, respectively. Thus, it gives rise to 
the most general usual perturbed FLRW metric

\begin{equation}
\begin{array}{r@{}l}
g_{00} = 1 + 2\psi,  \\ 
g_{i0} = a(\partial_i F + G_i), \\
g_{ij} = - a^2[(1-2\phi) + h_{ij} + \partial_i \partial_j B + \partial_j C_i \partial_i C_j].
\end{array}
\end{equation}

Here, we just treat the scalar perturbations with scalar modes $\psi$ and $\phi$ in the conformal Newtonian gauge 
as given in eq. (\ref{metric_N}).
The complete set of equations for scalar perturbations modes in the conformal Newtonian gauge for $f(T)$ gravity can be recast as 
(see \cite{LSS_fT})

\begin{equation}
\label{00}
 f_T k^2 \phi + 3 \mathcal{H}(\phi' + \mathcal{H} \psi) \Big[ f_T - 12 \frac{f_{TT}}{a^2}  \mathcal{H}^2 \Big] - 12 \frac{f_{TT}}{a^2}  \mathcal{H}^3 k \zeta = 4 \pi G a^2 \delta \rho,
\end{equation}

\begin{equation}
\label{0i}
 f_T k^2 (\phi' +  \mathcal{H} \psi) - 12 \frac{f_{TT}}{a^2} k  \mathcal{H} ( \mathcal{H}' -  \mathcal{H}^2) = 4 \pi G a^2 (\rho + p) \theta,
\end{equation}

\begin{eqnarray}
\label{ij1}
  f_T \Big[\phi'' +  \mathcal{H}(\psi' + 2 \phi') + (2 \mathcal{H}' +  \mathcal{H}^2)\psi + \frac{1}{3}k^2(\phi - \psi) \Big] - 48 \frac{f_{TTT}}{a^4} k  \mathcal{H}^3 ( \mathcal{H}' -  \mathcal{H}^2) \zeta \nonumber \\ 
 - 12 \frac{f_{TT}}{a^2} \Big[H^2\phi'' + H(3H'- H^2)\phi' + H^3\phi' + (5H' - 2H^2)H^2\psi \Big] \nonumber \\ 
 + 144 \frac{f_{TTT}}{a^4} \mathcal{H}^3( \mathcal{H}'- \mathcal{H}^2)(\phi' +  \mathcal{H} \psi) + 4 \frac{f_{TT}}{a^2} \Big[k^2  \mathcal{H} \zeta + k \mathcal{H} (3 \mathcal{H}' -  \mathcal{H}^2)\zeta  \Big] \nonumber \\ 
 = \frac{4 \pi}{3} G a^2 \delta p,
\end{eqnarray}

\begin{equation}
\label{ij2}
 f_T k^2 (\phi - \psi) + 12 \frac{f_{TT}}{a^2} k  \mathcal{H} ( \mathcal{H}' -  \mathcal{H}^2) \zeta = 12 \pi G a^2 (\rho + p) \sigma.
\end{equation}

In eqs. (\ref{00}) - (\ref{ij2}), $\mathcal{H}$ is the Hubble function defined with respect to the conformal time 
and prime denote the derivatives with respect to the conformal time.



In most modified gravity theories, the extra degree of freedom will be governed by a dynamical equation, for instance 
in $f(R)$ gravity \cite{fR_pertubation}. Here, in the above equantions, $\zeta$ is a constraint equation and takes the form

\begin{equation}
\label{zeta}
 \zeta = - \frac{3}{k} \Big(\frac{ \mathcal{H}' -  \mathcal{H}^2}{ \mathcal{H}} \phi + \phi' +  \mathcal{H} \psi \Big),
\end{equation}
which can be eliminated by a direct replacement of eq. (\ref{zeta}) in the above modified equations.

For a complete discussion of the above equations see \cite{LSS_fT}. For $f_T - 1 = f_{TT} = f_{TTT} = 0$, the equations (\ref{00}) - (\ref{ij2}) 
reduce to general relativity \cite{Ma_Bertschinger}. The expressions on the right-hand side of the equations (\ref{00}) - (\ref{ij1}) are respectively,

\begin{equation}
 \delta \rho = \delta \rho_{r} + \delta \rho_{m},
\end{equation}

\begin{equation}
 (\rho + p) \theta = (4/3) \rho_r + \rho_{m} \theta_{m}, 
\end{equation}

\begin{equation}
 \delta p = \delta p_r + \delta p_{m} = 1/3 \delta \rho_r,
\end{equation}
where the subscripts $m$ and $r$ represent cold dark matter plus baryons and photon plus neutrinos, respectively.
All species, baryons, dark matter, photons and neutrinos, do not undergo changes in their dynamics,
so the perturbation equations for each component follow the standard evolution as described \cite{Ma_Bertschinger}.

In the above quantities, $\theta_{i}$, with $i = r, m$, is the divergence of the fluid velocity.
In eq. (\ref{ij2}), the anisotropic stress perturbation $\Pi$ is connected to $\sigma$ via the relation $\sigma = 2 \Pi p/ 3(\rho + p)$. 

In general terms, relativistic particles respond to both potentials and non-relativistic particles respond to the time component of the metric, that is, 
to the potential $\psi$. For instance, neutrinos develop anisotropic stress after neutrino decoupling. Thus, $\phi$ and $\psi$
actually differ from each other in the time between neutrino decoupling up to matter-radiation equality.
After the universe becomes matter-dominated, $\phi$ and $\psi$ rapidly approach to each other (in the context of general relativity). 
The same happens to photons after decoupling, but the universe is then already matter-dominated, 
so the photons do not cause a signicant $\phi - \psi$ diference. So, in the context of general relativity, at late times, 
we expect $\phi = \psi$.  A gravitational slip, defined as $\phi \neq \psi$, generically occurs in modified gravity theories.
The gravitational slip is typically defined as

\begin{figure*}
 	\includegraphics[width=3.0in, height=2.5in]{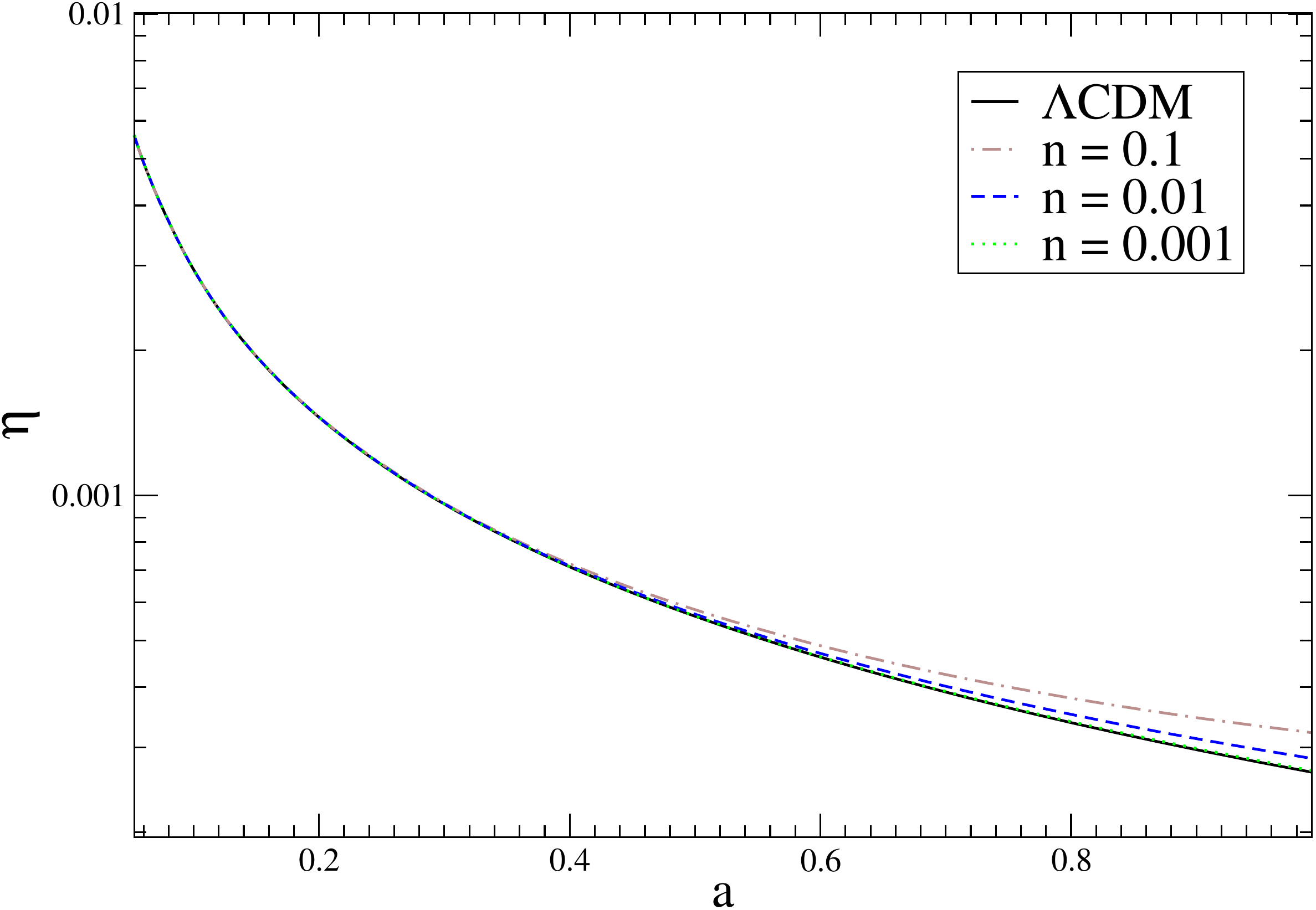} \quad \quad
 	\includegraphics[width=3.0in, height=2.5in]{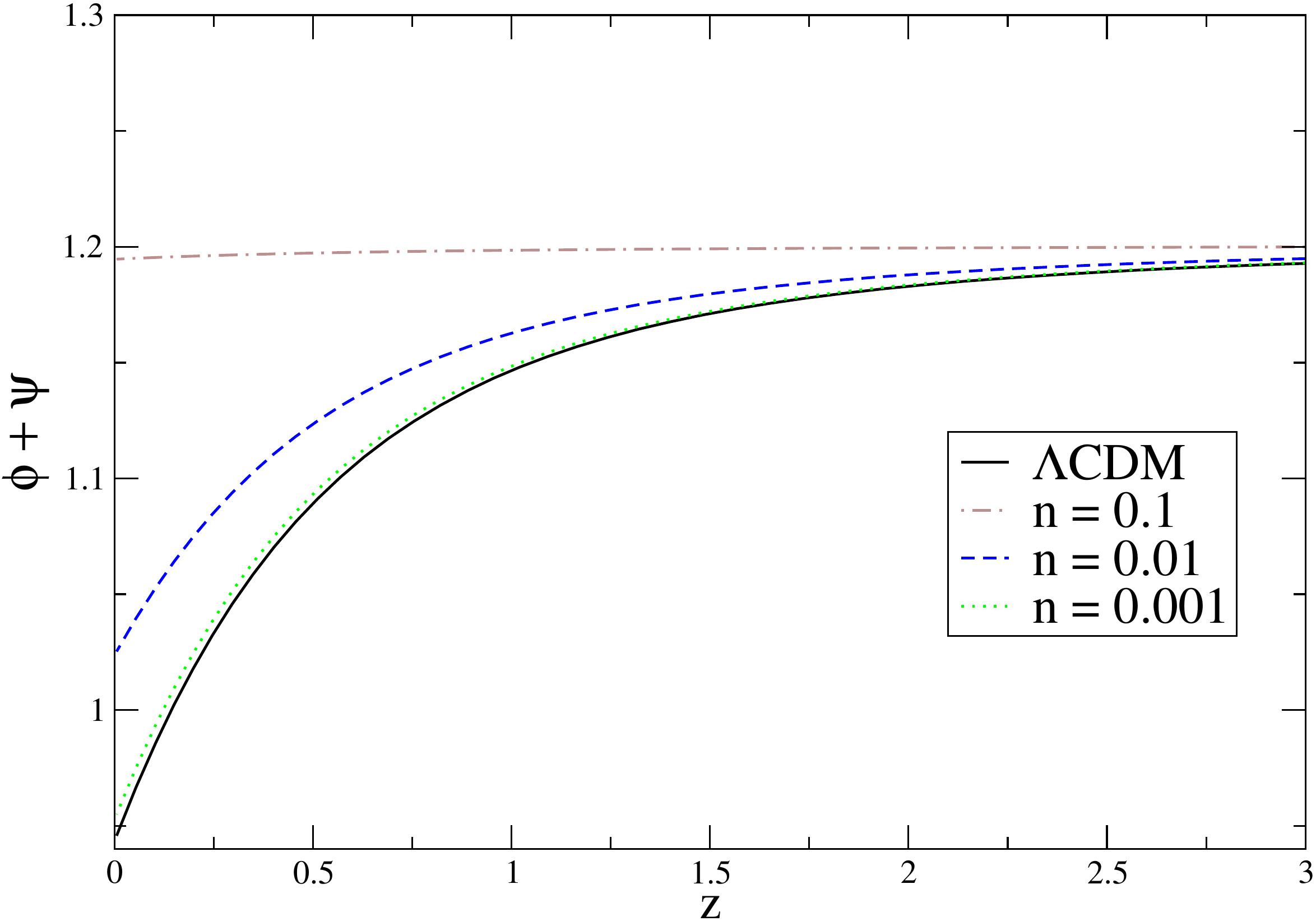}
 	\caption{Left painel: Evolution of the gravitational slip as a function of the scale factor $a$ to $k = 10^{-3}/{\rm Mpc}$ 
 	for the flat $\Lambda$CDM cosmology and $f(T)$ gravity for some values of $n$ from the model eq.(\ref{modf1}). 
 	Right panel: Evolution of the quantity $\phi + \psi$ at late time.}
 \label{potencial}
\end{figure*}

\begin{equation}
\label{eta}
 \eta = \frac{\phi-\psi}{\psi},
\end{equation}
where for $\eta \simeq 0$, there is no gravitational slip, that is, we have $\Lambda$CDM model. Let us quantify the behavior of 
the $\eta$ function in the next section.

The Fourier modes relevant to the linear regime of structure formation correspond approximately $k/aH \gg 1$, and remain well inside the horizon
in the redshift range $z < 2$. With that and also taking the quasi-static approximation \cite{Alessandra01}, where the time-derivatives 
of potential are very small, by combining equations (\ref{00}) and (\ref{0i}), in this approximation, we find

\begin{equation}
\label{Poisson_fT}
 k^2 \phi = 4 \pi G_{\rm eff} a^2 \delta \rho,
\end{equation}
which is the modified Poisson equation. Where we have defined 

\begin{equation}
\label{G_eff}
 G_{\rm eff} = \frac{G}{f_T},
\end{equation}
or more explicitly

\begin{equation}
 \frac{G_{\rm eff}}{G} = \frac{1}{1 + F_T}.
\end{equation}

Evidently for $F_T = 0$, one may recover the standard Poisson equation. The eq. (\ref{Poisson_fT}) is a standard parametrization for modified 
theories of gravity \cite{Alessandra01, Rachel} and this result here records a modification in the context of modified teleparallel gravity. The 
effective gravitational constant (\ref{G_eff}), has also been obtained in \cite{LSS_fT01}.

It will be interesting to investigate this modified Poisson equation using the data from redshift space distortion, 
where such modification may lead to interesting effects. 
The study of such effects would be an interesting avenue for future research. In what follows, let us 
quantitatively investigate the effects of the scalar perturbations generated via $f(T)$ gravity on CMB anisotropy.

\subsubsection{Effects on the CMB}

\begin{figure*}
 	\includegraphics[width=3.0in, height=2.5in]{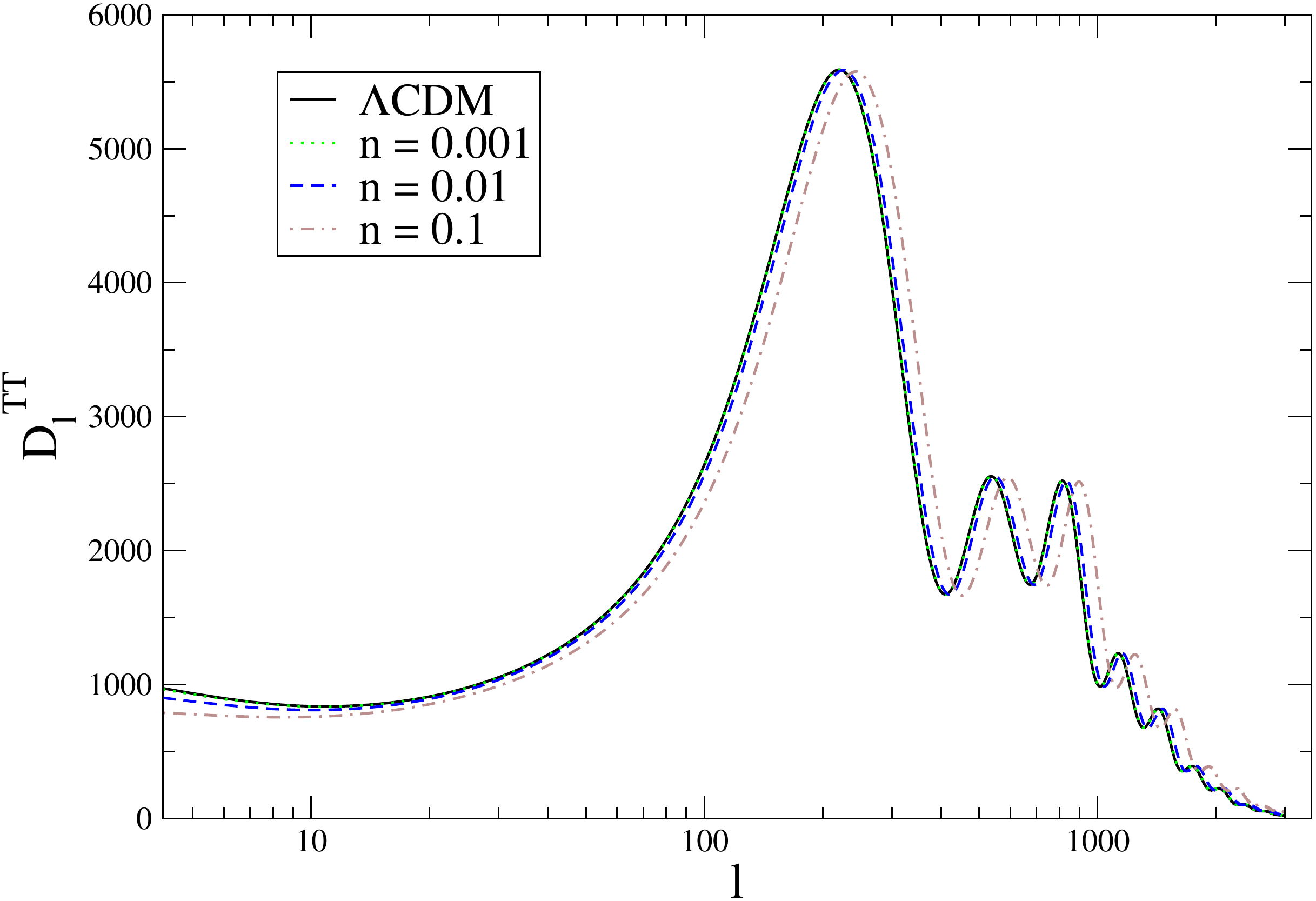} \quad \quad
 	\includegraphics[width=3.0in, height=2.5in]{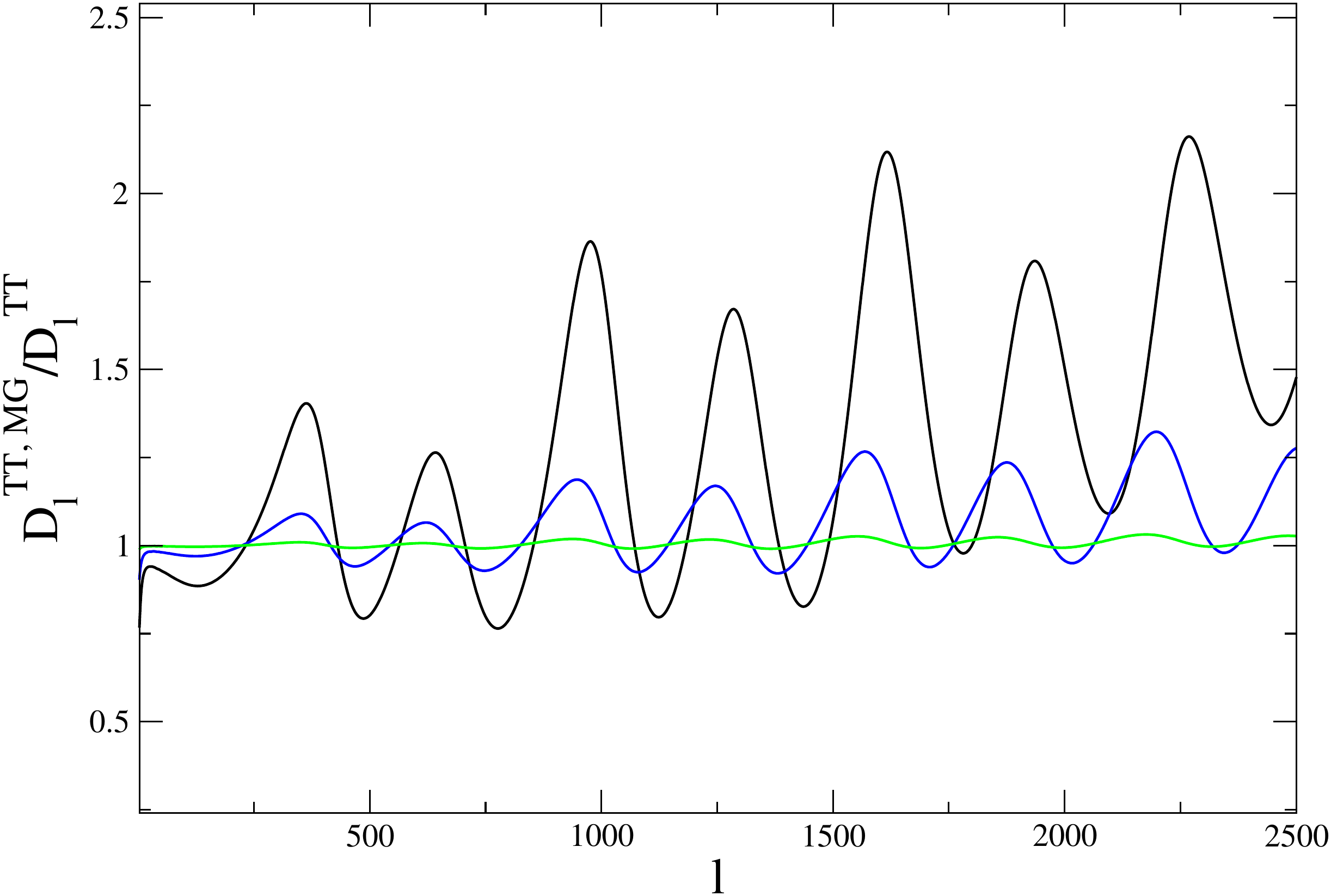}
 	\caption{Left panel: the CMB TT power spectrum, $D^{TT}_l =l(l+ 1)C_l^{TT}/ 2 \pi\mu K^2$, for the flat $\Lambda$CDM cosmology and $f(T)$ gravity 
 	 for some values of $n$ from the model eq.(\ref{modf1}). Right panel: Relative deviation of CMB TT power spectrum from the base line Planck 2015 
 	 $\Lambda$CDM model in comparison with $f(T)$ gravity for $n = 0.1$ (black line), $n = 0.01$ (blue line), and $n = 0.001$ (green line).}
 \label{cmb_TT}
\end{figure*}

\begin{figure*}
 	\includegraphics[width=3.0in, height=2.5in]{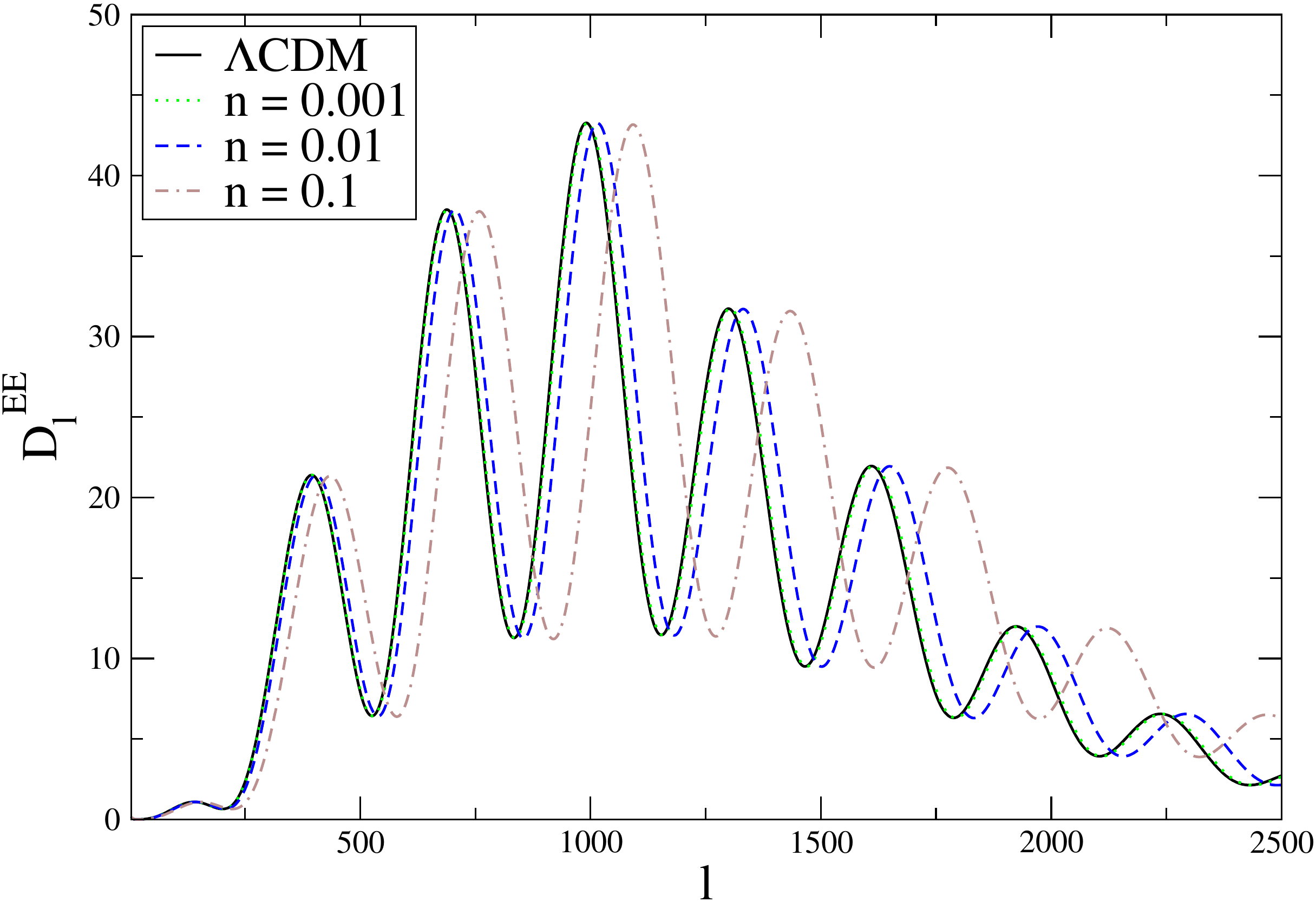} \quad \quad
 	\includegraphics[width=3.0in, height=2.5in]{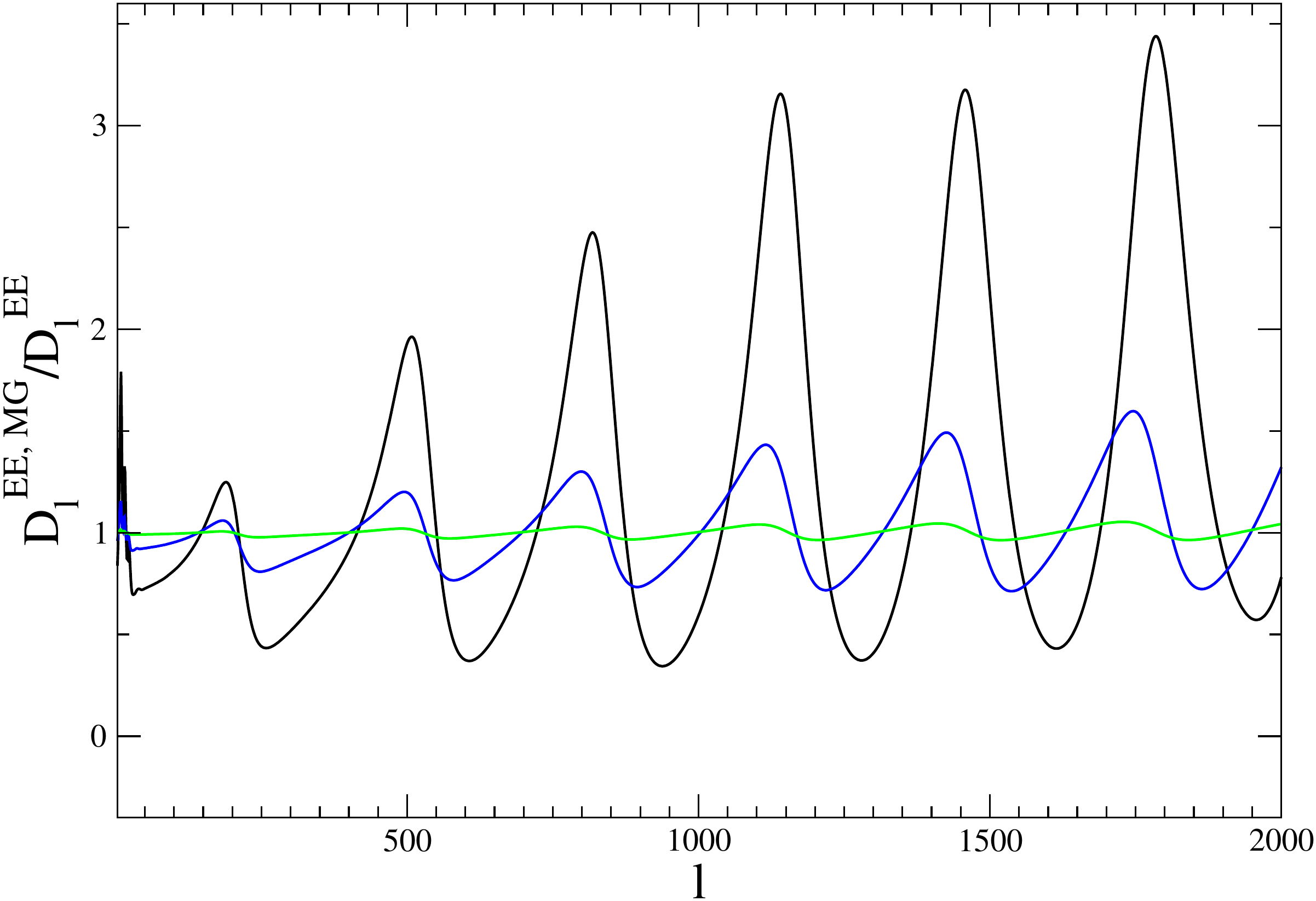}
 	\caption{The same as in figure \ref{cmb_TT}, but for the CMB EE power spectrum, $D^{EE}_l =l(l+ 1)C^{EE}_l/2 \pi\mu K^2$.}
 \label{cmb_EE}
\end{figure*}

In this section, in a nutshell, we describe the CMB temperature anisotropy, in order to see how $f(T)$ gravity can change this observable.
We know that the Newtonian curvature $\phi$ and potential $\psi$ encapsulate all observable properties of scalar fluctuations.
The contribution of a given $k$-mode to the amplitude of $l$th multipole moment of the CMB anisotropy is given by

\begin{equation}
\label{theta1}
 \Delta_l(\tau_0,k) = \int_{\tau_i}^{\tau_0} d\tau S_T(\tau, k) j_l [k(\tau_0 - \tau)],
\end{equation}
where the term source is given by \cite{Lesgourgues01, Hu01, Hu02}

\begin{equation}
\label{fonte}
 S_T(\tau, k) = g (\Theta_0 + \psi) + k^{-2}(g \theta_b)' + e^{-\kappa} (\phi' + \psi') + {\rm polarisation}.
\end{equation}

Assuming a random phase assumption for the $k$-modes, the scalar contribution to power spectrum of the anisotropies can be estimated as

\begin{equation}
\label{cls}
 C_l^{X Y} = 4 \pi \int \frac{dk}{k}  \Delta^X_l \Delta^Y_l P(k),
\end{equation}
where $X$ referring to the modes $T$, $E$, $B$ and $P(k)$ is the scalar primordial spectrum.

In eq. (\ref{fonte}), $g$ is the optical depth to Compton scattering between the present, $\tau_0$, and the epoch in
question $\tau$. The terms $\Theta_0$ and $v_b$ are the photon temperature perturbation and baryon velocity in Newtonian gauge, respectively. 
The terms, $\Theta_0 + \psi$ and $v_b$, are the ordinary Sachs-Wolfe and Doppler effect, respectively. They are responsible for the so-called
acoustic peaks in the CMB temperature spectrum. The term $\phi' + \psi'$ leads to the integrated Sachs-Wolfe effect and contributes after 
last surface scattering and leads to an additional, late time contribution to anisotropies in the CMB at large scales due to the decay 
of gravitational potentials in the presence of dark energy (or an effective dark energy via modified gravity).

We can note that the scalar perturbations in $f(T)$ gravity, described in the previous section, 
lead to a non-trivial change on the the dynamics of $\phi$ and $\psi$. Thus, it is expected that the effects of a given function $f(T)$ cause 
significant changes on CMB anisotropy in all angular scales. In order to quantify these effects, we introduce the simplest power-law model 
\cite{Bengochea:2008gz} given by

\begin{equation}
\label{modf1}
f(T)= T + \alpha (-T)^{n},
\end{equation}
where $\alpha$ and $n$ are two parameters of the model. Inserting this $f(T)$ form into  Friedmann equation (\ref{background1}) at present, 
we acquire

\begin{eqnarray}
\alpha=(6H_0^2)^{1-n}\frac{\Omega_{F0}}{2n-1},
\end{eqnarray}
where for $n=0$, the present scenario reduces to the $\Lambda$ + general relativity. 

It is well known that stability conditions play an important role in modified gravity models \cite{Raveri, Salvatelli, Peirone}, 
being able to generate strong impact when constraining the parameters of the theory and in selecting the priors \cite{Raveri, Salvatelli, Peirone}. 
In what follows, we only analyze the scalar perturbation of the model. But, it is known that in modified teleparallel gravity 
the speed of propagation for tensor mode is equal that in general relativity. At first, 
the running of the effective Planck mass ($\alpha_M$) is a function of $T$ (within an analogy with effective field theory approach), 
and so there may be a dependency between $\alpha_M$ and the free parameters of the theory, see \cite{Gw01} for details. 
Here, we only investigate the scalar modes evolution and as argued in \cite{Chen:2010va}, in this case the stability condition is satisfied 
in the range $0 < n < 1$ for the power-law model (\ref{modf1}). Also, another condition on the free parameter in eq. (\ref{modf1}) is that $n < 1$ 
in order to obtain an accelerating expansion of the Universe at late times.
As we shall see below, in order to explain CMB data, this free parameter needs to be some order of magnitude lower that 1. 
The statistical prior range of the baseline of the model will be specified in the next section.



We modified the publicly available CLASS code \cite{class} in order to calculate the evolution of the potential $\psi$ and $\phi$ 
and the CMB power spectrum, from the equations presented in the previous section 
(more specifically a direct implementation of the eqs. (\ref{0i}) and (\ref{ij2}) in perturbative sector)).

Figure \ref{potencial} on the left painel, shows the function $\eta$, eq. (\ref{eta}), as a function of the scalar factor $a$ 
for different values of $n$. As expected, the gravitational slip decay to $\eta \simeq 0 $ at late time in all cases, 
even considering a strong regime of scalar torsion ($n = 0.1$). The value $n = 0.1$ is considered a ``strong regime'', because as 
we will see in the next section, this value is some order of magnitude larger than the observational value of $n$. We note
numerically, at present time and $k = 10^{-3}/{\rm Mpc}$, $\eta = 3.21 \times 10^{-4}$, $\eta = 2.83 \times 10^{-4}$ and $\eta = 2.67 \times 10^{-4}$ 
for $n = 0.1$, $0.01$, and $0.001$, respectively. So, in general, the gravitational slip is of the same order of magnitude 
in all cases, but can slightly differentiate at present time. For $a < 0.3$, there is no difference.
Figure \ref{potencial} on the right panel, shows the function $\phi + \psi$ at late time. As expected, small
corrections like $n = 0.001$ have behavior very close (but different in numerical values) to the $\Lambda$CDM model. 
Significant variations can be easily noticed when the value of $n$ is increased. In the strong regime, 
the amount $\phi + \psi$ presented very small variations in relation to redshift. So, with the larger values of $n$,
the function $\phi + \psi$ tends to show small variations, or even remain constant. In practical terms, this can 
lead to direct consequence on CMB at large angular scales and significant effects on gravitational lensing, 
where the variation of the potencial plays an important role.

Figure \ref{cmb_TT}, on the left panel,  depicts the theoretical predictions for the CMB temperature power spectrum for $f(T)$ gravity as well as for 
the $\Lambda$CDM scenario. In order to have a better understanding of the 
effects due the modified teleparallel gravity, on the right panel of Figure \ref{cmb_TT}, we show the ratio between $f(T)$ gravity/$\Lambda$CDM.
For $n = 0.1$, we can see strong effects on all angular scales, as expected, 
once this value represents a strong cosmological regime of torsional scalar for all scales and cosmic time. 
On intermediary and small angular scales ($l > 400$) the difference with respect to $\Lambda$CDM becomes even greater. Thus, we can easily
summarize that values of this order of magnitude can already be ruled out, even theoretically.
Still for $n = 0.01$, we can notice significant changes, and only for $n = 0.001$ we have variations around
the $\Lambda$CDM model. Thus, it is expected that the free parameter that characterizes the 
modified teleparallel gravity models should be of the order of $10^{-3}$ or even smaller than it. 
For values greater than this order of magnitude, the evolution of the perturbations is well behaved, 
but are noticed very huge effects on the observables. For instance, making $n > 0.1$ completely incompatible with the observational data.

Figure \ref{cmb_EE} shows the theoretical predictions for the CMB temperature EE power spectrum. The pattern of the effects/deviations of the $f(T)$ gravity from that of 
the $\Lambda$CDM model on EE power spectrum is similar to the one observed in case of the CMB TT spectrum.

Without loss of generality, the quantified effects presented here are also expected to occur for any viable functional form of the $f(T)$ model
present in the literature, as for instance the exponential model. In what follows, we consider the 
parameterization (\ref{modf1}) in order to derive the observational constraints on $f(T)$ gravity for the first time 
using the full CMB temperature anisotropies data.  

\section{Observational constraints}
\label{results}

We modified the publicly available CLASS code \cite{class} together with Monte Python \cite{montepython} code for the $f(T)$ gravity 
scenario presented here. In order to constrain the model under consideration, we used the following data sets:
\\

\textbf{CMB}: The CMB data from Planck 2015 comprised of the likelihoods 
at multipoles $l \geq 30$ using TT, TE and EE power spectra and the low-multipole polarization likelihood at $l \leq 29$. 
The combination of these data is referred to as Planck TT, TE, EE + lowP in \cite{Planck2015}. We also include Planck 2015 CMB lensing 
power spectrum likelihood. 
\\

\textbf{BAO}: The baryon acoustic oscillations (BAO) measurements from the  Six  Degree  Field  Galaxy  Survey  (6dF) \cite{bao1}, 
the  Main  Galaxy  Sample  of  Data  Release 7  of  Sloan  Digital  Sky  Survey  (SDSS-MGS) \cite{bao2}, 
the  LOWZ  and  CMASS  galaxy  samples  of  the Baryon  Oscillation  Spectroscopic  Survey  (BOSS-LOWZ  and  BOSS-CMASS,  
respectively) \cite{bao3},  and the distribution of the LymanForest in BOSS (BOSS-Ly) \cite{bao4}. 
These data points are summarized in table I of \cite{baotot}.
\\
 
\textbf{H0}: We use the recently measured new local value of Hubble constant given by $H_0=73.24 \pm 1.74$  km s${}^{-1}$ Mpc${}^{-1}$ 
as reported in \cite{riess}.
\\

We used Metropolis Hastings algorithm with uniform priors on the model parameters to obtain correlated Markov Chain Monte Carlo samples by 
considering two combinations of data sets: CMB + BAO and CMB + BAO + $H_0$. 
We chose the minimal data set CMB + BAO since adding BAO data to CMB does not shift the regions of probability much, but
this combination constrains the matter density very well, and reduces the error-bars on the parameters. 
The other data set combination (CMB + BAO + $H_0$)  is considered in order
to investigate the effects of $H_0$ prior on the minimal data set. The baseline parameters set is

\begin{equation}
 \{ 100\omega_b, \, \omega_{cdm}, \, 100 \theta_s, \, \ln 10^{10} A_s, \, n_s, \, \tau_{reio}, \, n \},
\end{equation}
where the parameters correspond to baryon density, cold dark matter density, angular size of sound horizon at last scattering, 
the amplitude of initial scalar power spectrum, spectral index, the optical depth associated with reionization, 
the Hubble constant and $f(T)$ gravity free parameter, respectively. 
The priors used for the model parameters are: $\omega_b\in[0.005,0.1]$, $\omega_{cdm}\in[0.01,0.99]$, $100\theta_{s}\in[0.5,10]$, 
$\ln(10^{10}A_s)\in[2.4,4]$, $n_s\in[0.5,1.5]$, $\tau_{\rm reio}\in[0.01,0.8]$ and $n \in[0.0, 0.1]$.


\begin{table}[ht]
\label{tab1}
\begin{center}
\begin{tabular}{ccccc}

\hline
Parameter &  CMB + BAO  & CMB + BAO + $H_0$\\
\hline

{$10^{2} \omega_{b}$}  & $2.235_{-0.013}^{+0.013}$               & $2.235_{-0.013}^{+0.013}$       \\

{$\omega_{cdm }  $}   & $0.1181_{-0.001}^{+0.001}$             &   $0.118_{-0.001}^{+0.001}$     \\

{$100\theta_{s }  $}  & $1.041_{-0.00027}^{+0.00027}$            &  $1.041_{-0.00027}^{+0.00030}$      \\
 
{$\ln10^{10}A_{s }$}  & $3.078_{-0.023}^{+0.023}$              &  $3.08_{-0.022}^{+0.022}$      \\

{$n_{s}         $}   & $0.9678_{-0.0039}^{+0.0039}$              &  $0.9684_{-0.0044}^{+0.0039}$     \\

{$\tau_{reio }   $}  &  $0.073_{-0.013}^{+0.012}$              &    $0.075_{-0.012}^{+0.012}$    \\

{$n              $}  & $0.0043_{-0.0039}^{+0.0033}$     &    $0.0054_{-0.0020}^{+0.0020}$    \\

{$\log \alpha    $}  & $10.00_{-0.12}^{+0.081}$     &    $10.03_{-0.06}^{+0.06}$    \\

{$\Omega_{F0}    $}  & $0.73_{-0.028}^{+0.021}$     &    $0.738_{-0.015}^{+0.015}$    \\

{$H_0             $} & $72.4_{-4.1}^{+3.3} $            &   $73.5_{-2.1}^{+2.1}$     \\

{$\sigma_8         $} & $0.855_{-0.033}^{+0.023} $               &   $0.866_{-0.02}^{+0.02}$    \\

\hline
{$\chi^2_{min}/2 $}  & $6480.48$                           &    $6482.27$  

\end{tabular}\caption{\label{tab1} Constraints on the parameters from $f(T)$ gravity. Mean values of the parameters are displayed at 68\% C. L. 
The parameter $H_0$ is in the units of km s${}^{-1}$ Mpc${}^{-1}$.}
 \end{center}
\end{table}

\begin{figure}
\centering
\includegraphics[width=4.0in, height=3.0in]{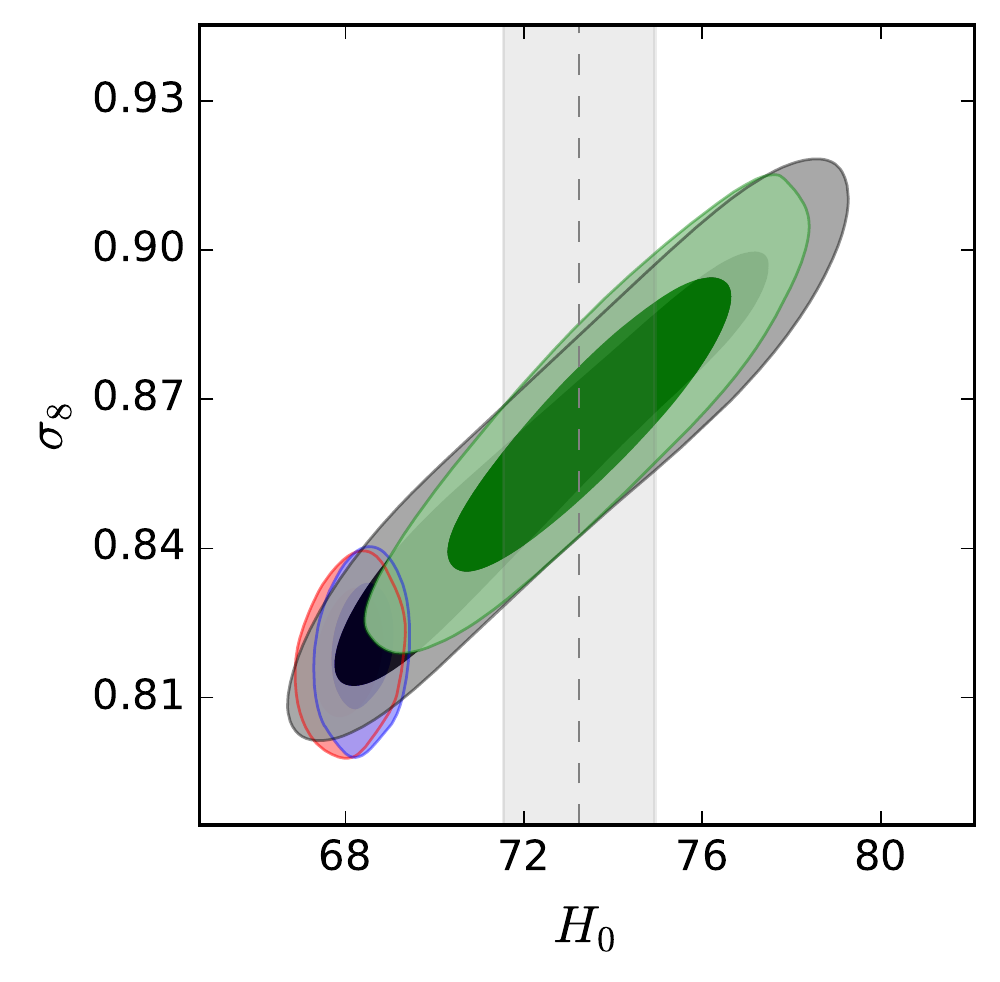}
 	\caption{Parametric space in the plane $H_0$ - $\sigma_8$, where the regions in red (blue) show the constraints for $\Lambda$CDM model from
 	CMB + BAO (CMB + BAO + $H_0$), respectively. The regions in black (green) show the constraints for $f(T)$ gravity from CMB + BAO
 	(CMB + BAO + $H_0$), respectively. The vertical gray band corresponds to $H_0=73.24\pm 1.74$ km s${}^{-1}$ Mpc${}^{-1}$.}
 \label{ps1}
\end{figure}

\begin{figure}
\centering
\includegraphics[width=4.0in, height=3.0in]{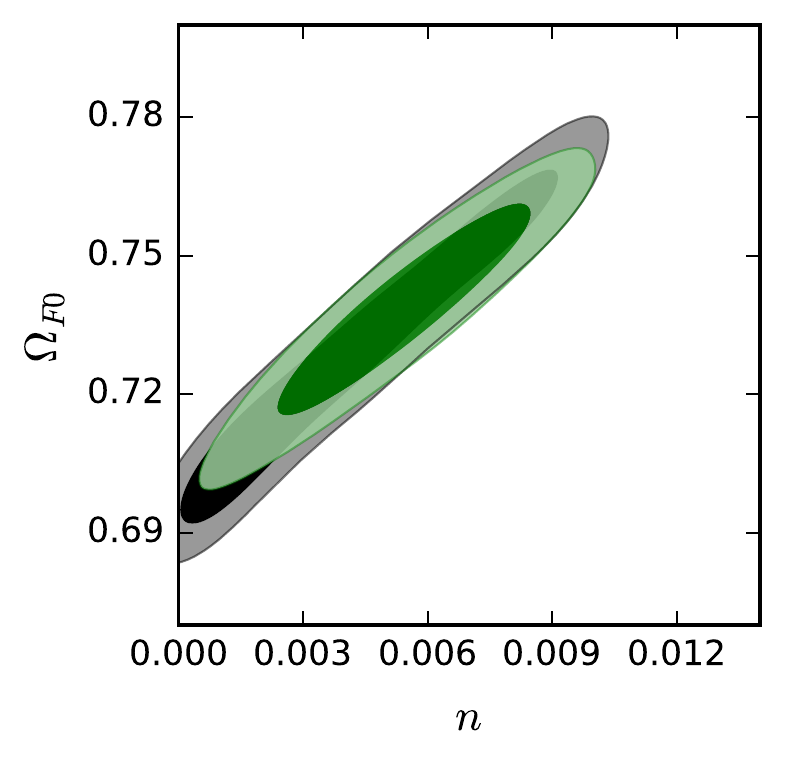}
 	\caption{Parametric space in the plane $n$ - $\Omega_{F0}$, where the regions in black (green) show the constraints from
 	CMB + BAO (CMB + BAO + $H_0$), respectively.}
 \label{ps2}
\end{figure}

Table \ref{tab1} summarizes the main results of the statistical analysis. Figure \ref{ps1} shows the parametric space at 68\% and 95\% 
confidence level (CL) in the plane $H_0$ - $\sigma_8$, both $\Lambda$CDM and $f(T)$ cosmology. The vertical gray band corresponds 
to $H_0=73.24\pm 1.74$ km s${}^{-1}$ Mpc${}^{-1}$ \cite{riess}.

Assuming standard $\Lambda$CDM model, the Planck team obtained $H_0 = 66.93 \pm 0.62$ km s${}^{-1}$ Mpc${}^{-1}$ that is about
3$\sigma$ CL deviations away from the direct and local determined value $H_0=73.24 \pm 1.74$ km s${}^{-1}$ Mpc${}^{-1}$ as
reported in \cite{riess}. We can note that $f(T)$ gravity gives a higher value of $H_0$ (see figure \ref{ps1}). Therefore, 
the discrepancy between the $H_0$ values can be removed here. We have found 
$H_0 = 72.4_{-4.1}^{+3.3}$ km s${}^{-1}$ Mpc${}^{-1}$ using CMB + BAO data alone. This constraint is fully compatible with the local determination 
of $H_0$ \cite{riess}. Evidently, the combination of the minimal data set with $H_0$, that is, using a prior on $H_0$ in the analysis, also 
yields the $H_0$ value compatible with the local measurement. Other physical models beyond the minimal $\Lambda$CDM model have also been 
considered to solve the tension on $H_0$ \cite{H01,H02,H03,H04,H05,H06,H07,H08}.

Another cosmological tension arises from the predictions of the direct measurements of LSS and CMB for $\sigma_8$.
The results from Planck CMB yield the value of amplitude of matter density fluctuations, $\sigma_8= 0.831\pm 0.013$ \cite{Planck2015},
which is about $2\sigma$ higher than $\sigma_8= 0.75\pm 0.03$ as given by the Sunyaev-Zel’dovich cluster abundances measurements \cite{p14}, 
for example. From our constraints, we note that the model is not able to solve the tension $\sigma_8$. In order to have a model, within 
modified teleparallel gravity context, capable of solving both tensions at the same time, an extension
with a detailed investigation, beyond the model presented here, should be done.

The dynamics generated by modified teleparallel gravity is able to solve the tension on $H_0$ and as we can see from
figure \ref{ps1}, the model provides wide range of values for $H_0$. 
Figure \ref{ps2} shows the parametric space at 68\% and 95\% CL in the plan $n$ - $\Omega_{F0}$. It is important to remember that $n$ is the only 
free parameter of the model and only this parameter has effects on the behavior of the model. 
Table \ref{tab1} also shows the constrains on the derived parameters $\alpha$ (given in terms of the its absolute value and in logarithmic scale) 
and $\Omega_{F0}$. We can note from CMB + BAO analysis that the model is compatible with $\Lambda$CDM ($n =0$). Analyzing CMB + BAO + $H_0$, 
the parameter $n$ can deviate from $\Lambda$CDM up to approximately 2.2$\sigma$ CL.

Assuming $\Lambda$CDM we have, $\chi^2_{min}/2 = 6480.91$ ($\chi^2_{min}/2 = 6486.33$) from CMB + BAO (CMB + BAO + $H_0$), respectively.
Thus, defining $\Delta \chi^2 = \chi^{2, \, \Lambda CDM}_{min} - \chi^{2, \, f(T)}_{min}$, we find $\Delta \chi^2 = 0.43$ and 
$\Delta \chi^2 = 4.06$ from CMB + BAO (CMB + BAO + $H_0$), respectively. We see that statistically both
analyzes yield an improvement over $\Lambda$CDM model. 
In case of CMB + BAO, the improvement is minimal while from CMB + BAO + $H_0$ it is around 2.2$\sigma$ CL.

Finally, we close the observational analysis section by comparing the statistical fit using the standard information criteria via Akaike 
Information Criterion (AIC) \cite{AIC1,AIC2}, $AIC = \chi^2_{min} + 2 d $, where $d$ is the number of free model parameters. In $f(T)$ gravity, we
have one free parameter more than in $\Lambda$CDM model. Assessing $\Delta AIC$ concerning to $\Lambda$CDM, we find $\Delta AIC = -1.57$ 
($\Delta AIC = 2.06$) from CMB + BAO (CMB + BAO + $H_0$), respectively.  The thumb rule of AIC reads that the models are 
statistically indistinguishable from each other and a small negative (positive) evidence from CMB + BAO (CMB + BAO + $H_0$).

\section{Final remarks}
\label{Conclusions}

In the present work, we have investigated the effects of scalar linear perturbations in the context of modified teleparallel gravity
and analyzed the consequences on the CMB temperature power spectrum. It is expected that a very small correction in $f(T)$ gravity, 
beyond the $\Lambda CDM$, could be able to generate the fluctuations of temperature expected by the observations. 
Also, the first time, we report the observational constrains on $f(T)$ gravity using the  data from the CMB observations 
(Table \ref{tab1} summarizes the results). The observational data indicate that modified teleparallel gravity depicts a small deviation 
from the $\Lambda$CDM-cosmology up to approximately 2.2$\sigma$ CL . We find that the $f(T)$ gravity is consistent with CMB observations, 
and it can serve as a viable candidate in the class of modified gravity theories.  
An important result of this development is that the $f(T)$ gravity can naturally solve the $H_0$ tension between the local value of the Hubble
constant and the value predicted by the most recent CMB anisotropy data of the Planck satellite in $\Lambda$CDM-cosmology.
Evidently, the observational constraints can be extended to other parametric functions of the scalar torsion, 
as well as one may add other data sets in the analysis.  

In addition to the developments and results already known in the literature within the context of 
modified teleparallel gravity, the results presented in this work offer a complement and 
a step forward in this direction. On the other hand, it will be interesting to investigate how the tensor modes evolution, that is, 
gravitational waves in $f(T)$ gravity \cite{Gw01,Gw02} can influence CMB power spectrum, especially the B-modes polarization. Also, we proposed recently 
a nonlocal formulation of teleparallel gravity \cite{nonlocal01}. Therefore, the development of scalar perturbations
and investigating the large scale structure formation  need further attention in this sense.

\begin{acknowledgments}
\noindent  I am grateful to S. Kumar, S. Pan and E. N. Saridakis for the discussion and a critical reading of the manuscript. Also, 
I am thank the referee for his/her comments and suggestions.

\end{acknowledgments}

\end{document}